\documentclass[pdflatex,sn-mathphys-num]{sn-jnl}

\usepackage{graphicx}%
\usepackage{multirow}%
\usepackage{amsmath,amssymb,amsfonts}%
\usepackage{amsthm}%
\usepackage[mathscr]{eucal}
\usepackage[title]{appendix}%
\usepackage{xcolor}%
\usepackage{textcomp}%
\usepackage{manyfoot}%
\usepackage{booktabs}%
\usepackage{algorithm}%
\usepackage{algorithmicx}%
\usepackage{algpseudocode}%
\usepackage{listings}%

\usepackage{amsthm}
\usepackage{lmodern} 
\usepackage{amsmath} 
\usepackage{fix-cm}
\fontsize{6}{7}\selectfont

\UseRawInputEncoding

\usepackage{afterpage}

\begin{document}

\title{Transforming Teacher Education in Developing Countries: The Role of Generative AI in Bridging Theory and Practice}

\author{\fnm{Matthew} \sur{Nyaaba}}
\affil{\orgdiv{AI4STEM and Department of Educational Theory and Practice}, \orgname{University of Education}, \city{Athens}, \state{Georgia}, \country{USA}}
\email{\textbf{Author's Email:} matthew.nyaaba@uga.edu}

\abstract{This perspective study examines the potential roles and impacts of Generative AI (GenAI) in addressing persistent challenges in teacher education within developing countries, using Ghana as a case. A review of Ghana's teacher education reforms reveals longstanding challenges in implementing pedagogical modeling and performance-based assessments, overshadowed by the dominance of content-based teaching and standardized assessment approaches. Employing a logical deduction method, this study argues that GenAI could act as a catalyst for systemic change in addressing these challenges by alleviating the burden of content knowledge acquisition, thereby freeing up time for teacher educators to focus on pedagogical modeling and performance-based assessments. The effect will set a chain reaction in motion, addressing several other challenges in teacher education, such as curricular infidelity stemming from resistance to change among teacher educators, insufficient resources resulting from institutional or governmental neglect, overreliance on standardized testing, and gaps in practitioner expertise, etc. Consequently, this will equip pre-service teachers (PSTs) with relevant pedagogical skills, enabling them to enhance the learning outcomes of their future K-12 students. As a result, in the long term, this shift will bridge the gap between education and the skills required in the workforce, ultimately contributing to a reduction in unemployment. However, there should be caution against the misuse of GenAI, as it has the potential to even undermine existing critical thinking and creativity skills that are nurtured through the current traditional teaching methods, like lecture-based content delivery. Following these immediate and long-term impacts, this study calls for the urgent consideration and implementation of GenAI in teacher education through a scaffolded approach. The study further recommends empirical research to explore each role of GenAI and the practical steps necessary to ensure its responsible integration into teacher education.}




\maketitle

\section{Introduction}\label{sec1}

The rapid evolution of education systems globally demands that teacher education programs equip pre-service teachers (PSTs) with the skills and knowledge needed to foster inclusive, learner-centered environments \cite{Mishra2023}. However, despite extensive efforts to improve teacher education, many programs, especially in developing countries, still struggle to achieve the necessary balance between imparting content knowledge and providing practical, hands-on training \cite{EkizKiran2021, Pinamang2016}. Teacher education programs often struggle with inadequate commitment from implementers and government agencies, largely due to an over-reliance on content delivery that neglects critical components such as pedagogical modeling, practical application, and performance-based assessment. As a result, there is a growing disconnection between the theoretical underpinnings of teacher education and the practical realities of classroom instruction \cite{Nyaaba2021, Ogunniyi2015}.

However, recent research has shown that Generative AI (GenAI) is currently transforming teacher education in diverse ways \cite{Melek2024, Nyaaba2024Buddy}. As highlighted in recent studies, GenAI can serve as a tool for revitalizing teacher education in developing countries, providing PSTs with access to resources, supporting their content knowledge, and enhancing their teaching tasks \cite{Lee2024,  Nyaaba2024Buddy}. However, \cite{Sperling2024} scoping review on GenAI in teacher education indicates that there is significant lack of literature on GenAI and teacher education, particularly in developing countries. They further indicated that there is limited focus on how these GenAI tools can effectively enhance pedagogical practices, inform professional development, or address systemic barriers to effective implementation of teacher education curricula.

This study, therefore, presents a comprehensive exploration of the transformative role of Generative AI (GenAI) in teacher education, emphasizing its potential to create ripple effects across the broader education system and align with societal and workforce needs in developing countries. This study specifically examines Ghana`s teacher education reforms as a case, analyzing the challenges and successes within its numerous reform initiatives to achieve effective teacher preparation. The study finally highlights the key and invisible roles GenAI can play in transforming teacher education through the effective implementation of teacher education curricula in ongoing reforms.

\section{Recent Studies on Generative AI in Teacher Education}
Recent studies on the integration of GenAI in teacher education have explored its use, benefits, and challenges, particularly from the perspective of PSTs. \cite{Markos2024} investigated the perceptions of 257 PSTs in Greece, focusing on GenAI's strengths, weaknesses, opportunities, and threats. The findings revealed that while PSTs viewed GenAI as a valuable tool for self-directed learning and content support, concerns were raised about accuracy and the potential for over-reliance on GenAI tools. Similarly, \cite{Nyaaba2024Buddy} study in Ghana with 167 PSTs highlights the positive perceptions of GenAI among PSTs, who use it as a learning tool that improves their content knowledge, assists in lesson planning, and enhances access to teaching resources. Moreover, \cite{ LeeZhai2024} explored how 29 PSTs integrated ChatGPT into science lesson plans, identifying its diverse application across domains like Biology, Chemistry, and Earth Science. Although the lesson plans scored well on an AI-technological pedagogical and content knowledge (AI-TPACK) rubric, the study noted the underutilization of ChatGPT's full capabilities and improper use of AI-generated information.

In addition to student perceptions, teacher educators have shown increasing interest in integrating GenAI into their instructional practices. Another study by \cite{Nyaaba2024Professional} examined the professional development needs of teacher educators in Ghana, which revealed varying levels of familiarity with GenAI tools by teacher educators. The teacher educators acknowledged GenAI's potential in grading and generating classroom materials, but challenges such as high costs, limited training, and ethical concerns hindered widespread adoption. These studies collectively emphasize that GenAI has potential roles in transforming teacher education programs, particularly in developing countries.

\section {Initial Teacher Education Reforms in Ghana}
The reform of initial teacher education in Ghana has been a gradual yet significant process, marked by a shift from certificate and diploma programs to a more comprehensive four-year bachelor's degree \cite{Graham2013, Mpuangnan2020, TTEL2017}. This transformation aimed to address inadequacies in the previous teacher education systems, which were heavily content-driven with limited focus on practical pedagogical training. Under the former certificate programs, the emphasis was on rapidly preparing teachers to meet the growing demand for educators in basic schools \cite{AddaeKyeremeh2024}. However, the need for higher-quality education, more skilled teachers, and alignment with international standards led to the introduction of diploma programs and ultimately the four-year degree \cite{AddaeKyeremeh2024, Ntumi2023}. The bachelor's degree program was designed to offer a more holistic approach to teacher preparation by integrating both content and pedagogy with extended periods of teaching practice, ensuring that PSTs develop the skills necessary for effective classroom teaching in diverse settings \cite{Jiang2023}.

This transition was formalized through the passage of the Colleges of Education Act 2012 (Act 847), which elevated the status of Colleges of Education to tertiary institutions \cite{Buabeng2020, TTEL2017}. This legislative reform enabled these institutions to offer degree-level qualifications, marking a significant step forward from the previous diploma structure. The new four-year program combines content courses, pedagogical training, and comprehensive practicum experiences, including both on-campus peer teaching and off-campus field practice \cite{TTEL2017}. The primary objective is to ensure that PSTs graduate with both a deep understanding of subject matter and the practical skills needed to engage students effectively in modern classroom environments \cite{Mpuangnan2021}. However, challenges such as limited resources, resistance to change, and infrastructural deficits persist, impeding the full realization of the potential benefits of the four-year degree program \cite{Issaka2022, Kuyini2011}.

\subsection{Theoretical Framework of Initial Teacher Education in Ghana}
PThe theoretical framework underpinning the initial teacher training in Ghana is rooted in constructivist and experiential learning theories \cite{Bada2015, Kolb2014}. These frameworks emphasize the development of reflective practitioners who can adapt to diverse classroom situations \cite{Kolb2014}. Constructivism serves as the foundation for the curriculum, highlighting the importance of active learning, where students construct their own understanding of material rather than passively receiving information. This focus on learner-centered pedagogies encourages PSTs to engage critically with content, challenge assumptions, and create meaning through hands-on activities and real-world teaching experiences \cite{Shah2020}. Kolb et al.'s \cite{Kolb2014} theory of experiential learning further supports this approach by providing a structured model for learning through doing. The practicum and field placements embedded within the program give PSTs the opportunity to reflect on their experiences, refine their teaching practices based on feedback, and continuously improve their skills. Additionally, the integration of pedagogical content knowledge (PCK), as described by \cite{Hu2024, Pinamang2016}, ensures that teachers are not only experts in their subject areas but also skilled in delivering that content effectively in the classroom.

\subsection{Core and Transferable Skills in Initial Teacher Training in Ghana}
The four-year initial teacher education program in Ghana places significant emphasis on the development of core and transferable skills. These essential competencies include digital literacy, critical thinking, collaboration, communication, personal development, respect for diversity, and accountability \cite{AddaeKyeremeh2024, Mpuangnan2020}. Integrated across the curriculum, these skills are designed to equip PSTs with tools to address cross-cutting issues such as equity and inclusion. For instance, digital literacy is promoted by offering opportunities for PSTs to engage with technology, surf the internet, and present their findings using modern digital tools \cite{Education2017, Quaicoe2020}. This focus is crucial in ensuring that future educators can leverage technology to enhance learning in increasingly digital classroom environments.

Critical thinking is nurtured through reflective practices, data collection, and analysis, enabling PSTs to develop problem-solving abilities essential for navigating complex teaching challenges \cite{Asare2014, Schendel2023}. Collaboration is encouraged through group projects and presentations, fostering teamwork and building positive relationships between schools and communities \cite{Schendel2023}. Communication skills, considered a vital component of effective teaching, are honed through activities that require PSTs to present, examine, and challenge misconceptions, as well as articulate their teaching philosophies \cite{Asare2014, Schendel2023}.

The program also highlights personal development and inquiry skills, particularly through action research, where PSTs learn to gather and analyze data, then initiate interventions tailored to the needs of individual learners or small groups. Respect for diversity and individual differences is reinforced through reflective practices that encourage PSTs to design inclusive interventions. Moreover, honesty and accountability are ingrained in the program, requiring PSTs to collect accurate data on learners and present findings responsibly. Together, these core and transferable skills form the foundation for developing educators who are not only proficient in pedagogy but also capable of adapting to the demands of a diverse and dynamic teaching profession.

\section{Pedagogical Modeling}
Pedagogical modeling refers to the practice where teacher educators demonstrate effective teaching strategies and techniques to PSTs, showing how educational theories can be translated into real classroom practice \cite{Struyven2010, Swennen2008}. In developing countries like Ghana, this modeling is crucial as it equips future teachers with the skills and understanding needed to implement progressive teaching methods \cite{Campbell2015}. However, a significant challenge arises when teacher educators themselves rely on traditional, lecture-based approaches, despite promoting modern, student-centered strategies like multimedia integration and active learning \cite{Struyven2010}. This gap leaves PSTs with strong theoretical foundations but limited hands-on experience in applying these innovative methods in the classroom. Pedagogical modeling requires teacher educators to demonstrate how learner-centered approaches can be adapted to various teaching contexts, including resource-constrained environments \cite{Ponet2023}. The goal is to help PSTs to see the practical application of progressive teaching techniques, preparing them to create dynamic, engaging classrooms, which traditional methods might have undermined in improving educational outcomes and student engagement.

\section{Performance-based Assessment }
Performance-based tasks are integral to assessing students' abilities by engaging them in practical activities such as projects, presentations, or portfolios \cite{Buabeng2020, Elam1971}. These assessments provide a more authentic measure of students' knowledge and skills, as they allow for the demonstration of real-world application and critical thinking, beyond what standardized tests can offer \cite{Harris1997}. In many developing countries like Ghana, while performance-based assessment is acknowledged in the education system, its implementation remains limited, particularly in teacher education programs \cite{Asare2014, Mpuangnan2020}. The only widely respected form of authentic performance-based assessment in these programs is the practicum or supported teaching in schools (STS), where PSTs are assessed on their ability to teach in actual classroom environments \cite{AbdulKarim2023, Otoo2021, TTEL2017}. However, beyond this, other forms of performance-based assessments are often underutilized, despite their potential to enhance the practical readiness of future teachers.

Nonetheless, the three assessment components for each course largely emphasize performance-based assessments as follows: Continuous assessment; subject portfolio - 30\%, subject project - 30\%, Summative assessment, end of semester exam - 40\% \cite{NTC2017}. The current reform in teacher education in Ghana focuses more on developing practical skills and competencies that align with the core competencies outlined in the curriculum framework. Subject portfolios are used to collect and showcase student teachers' best work, demonstrating their growth and providing evidence of their developing skills in relation to Course Learning Outcomes (CLO) and National Teaching Standards (NTS). Subject projects, on the other hand, offer opportunities for student teachers to engage deeply in real-world problems, fostering independent learning, collaboration, and cross-disciplinary understanding \cite{NTC2017, TTEL2017}. Whereas the end-of-semester examinations assess both content and professional knowledge, ensuring standardized and equitable assessment practices across institutions \cite{NTC2017}. These comprehensive assessments aim to reduce the reliance on rote learning and enhance the quality of education delivered by future teachers \cite{Ntumi2023}.

\section{Challenges Facing Pedagogical Modeling in Teacher Education}
The implementation of pedagogical modeling in teacher education in developing countries faces substantial barriers that impede its effectiveness. While broader macro- and political systemic challenges complicate efforts to integrate innovative teaching practices, often leaving PSTs underprepared to adopt learner-centered pedagogies, institutional factors also play a crucial role. This section explores the key institutional-level challenges, focusing on stakeholders such as teacher educators, whose practices and attitudes significantly influence the effectiveness of curriculum implementation. 

\subsection{Inadequate Training and Experience in Innovative Pedagogies}
A critical issue in the effective implementation of pedagogical modeling is the lack of training and experience among teacher educators in innovative, learner-centered pedagogies \cite{DarlingHammond2016}. Many educators in Ghana were themselves trained under traditional, lecture-based systems that emphasized rote learning over interactive pedagogy \cite{Akyeampong2017b, Annan2020}. While reforms in Ghana have introduced new teacher education programs, transitioning from certificate to diploma and degree qualifications, these changes have not fully addressed the gaps in experiential training \cite{AntwiBoampong2018}. Most teacher educators continue to rely on outdated methods, limiting the opportunities for PSTs to observe or practice innovative, inclusive, and learner-centered teaching techniques \cite{Adarkwah2021, Nketsia2020}. As a result, the disconnect between theoretical concepts and practical classroom applications remains a significant challenge, despite reforms aimed at promoting pedagogical modeling. 

Furthermore, a significant challenge within teacher education in Ghana is that most teacher educators lack direct experience as practitioners in early childhood and basic education; instead, they are specialists in fields such as pure mathematics or science \cite{Akyeampong2017b}. This issue, compounded by both the scarcity of such practitioners and prevailing misconceptions about their recruitment by principals and hiring institutions, underscores the critical role of professional development (PD) sessions in teacher education programs in Ghana \cite{Akyeampong2017a, Sambo2020}. In light of this, as part of the implementation of Ghana's new curriculum, continuous PD sessions aimed at improving teaching practices have been introduced. However, the effectiveness of these sessions has often been limited due to factors such as time constraints, lack of resources, and the traditional focus on content delivery.

\subsection{High Course Loads and Limited Time}
Another barrier to pedagogical modeling is the overwhelming course loads and limited time available to teacher educators \cite{Gregory2015}. In Ghana, for instance, many teacher educators often manage large class sizes and are burdened with administrative duties such as marking and grading, research expectations, and community engagements \cite{Annan2020, Gregory2015}. These demands make it difficult for educators to dedicate the necessary time and effort to implement interactive, experiential teaching methods and performance-based assessments \cite{Asare2014}. The pressure to cover extensive material in a limited time often leads teacher educators to default to lecture-based instruction \cite{Buabeng2020}. Consequently, PSTs graduate with a strong theoretical foundation but limited hands-on experience in applying student-centered teaching practices, which eventually affects K-12 students.

\subsection{Resistance to Change and Institutional Culture}
Resistance to change is a significant obstacle in the adoption of pedagogical modeling \cite{Jost2015}. In many cases, both teacher educators and institutions are reluctant to fully embrace innovative, learner-centered approaches. In Ghana, this resistance can partly be attributed to infrastructure deficits and traditional attitudes toward education, which remain prevalent with a continued focus on content delivery and standardized examination results \cite{AntwiBoampong2018}. Teacher educators who have not received sufficient professional development in innovative teaching methods may feel uncomfortable with or resistant to adopting these practices. Institutional cultures that prioritize efficiency, research output, and content coverage over pedagogical innovation further compound this resistance \cite{Khosa2021}.

\subsection{Lack of Resources and Infrastructure}
The lack of access to resources and infrastructure is another challenge that affects the implementation of pedagogical modeling in Ghana. Many teacher education programs are under-resourced, lacking the necessary tools, materials, and technological infrastructure to support active, experiential learning \cite{Asare2014, Mpuangnan2020}. Digital resources such as smart boards or learning management systems are often absent, and the high cost of internet data limits opportunities for educators to create engaging, hands-on learning experiences \cite{Aboyinga2020, Acquah2019}. 

Pre-service teachers often lack sufficient practical experience, leaving them ill-prepared to engage students effectively or adapt their teaching strategies to diverse learning needs \cite{Gay2000, Grant2021}. In Ghana, these challenges are even more pronounced due to larger class sizes, resource limitations, and the complex cultural contexts within which teachers operate \cite{Akyeampong2017a, Buabeng2020, Pryor2012}. This contributes to the persistent reliance on traditional lecture-based methods that hinder the effectiveness of the numerous teacher education reforms aimed at addressing PSTs' lack of pedagogical modeling and performance-based assessment \cite{AddaeKyeremeh2024, Coffie2019}. 

There is an acknowledgment that broader macro and political systemic challenges complicate efforts to integrate innovative teaching practices. However, institutional-level challenges, particularly those related to stakeholders like teacher educators, remain crucial \cite{Buabeng2020, Nahal2010}.

\section{Role of Generative AI in Teacher Education}
\begin{figure}[H]
\centering
\includegraphics[width=\textwidth]{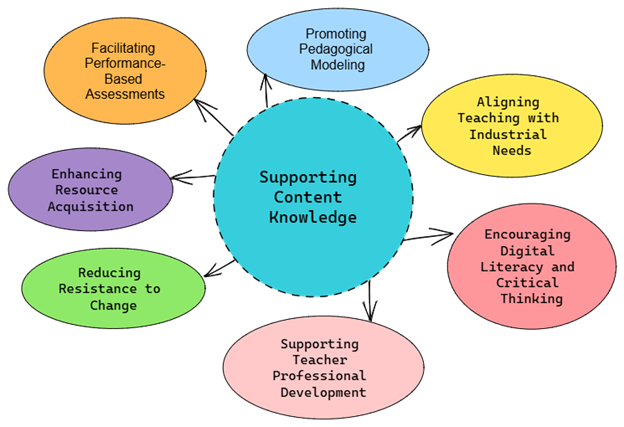} 
\caption{: GenAI Role in Transforming Teacher Education}
\label{fig:Picture1.png}
\end{figure}

This study argues that GenAI possesses the capacity to transform multiple facets of teacher education programs. At the core of the transformative role of GenAI is its potential to support content knowledge acquisition, which serves as the bedrock for enabling other critical roles in teacher education. By alleviating the time and effort spent on traditional content delivery, GenAI allows educators to focus on advanced pedagogical modeling, address resistance to change, and promote performance-based assessments. Furthermore, GenAI has the potential to enhance access to diverse resources, encourage digital literacy and critical thinking, align teaching with workforce demands, and support ongoing professional development for educators. This hierarchical interdependence, with content knowledge as the central pillar, underscores the pivotal role of GenAI both directly and indirectly, creating chain reactions across the entire teacher education program (See Fig.  \ref{fig:Picture1.png})

\subsection{Supporting Content Knowledge}
Research has shown that GenAI is highly capable of supporting content-based learning while facilitating personalized learning experiences. GenAI has the potential to synthesize content efficiently, presenting an opportunity to reduce reliance on static and restrictive materials. Pre-service teahers are increasingly turning to tools like ChatGPT to help with their content learning and teaching, making these tools valuable supplements for teacher educators to use in subject-specific areas \cite{SuYang2023}. For instance, a study by \cite{Kasneci2023} indicates that GenAI can effectively support PSTs in content knowledge acquisition and task development. Their research demonstrated that while GenAI-assisted PSTs performed similarly to those using traditional textbooks in terms of task correctness, they faced challenges with clarity and context embedding. This highlights the potential of GenAI tools to aid PSTs in developing content, though careful oversight is required to ensure high-quality outcomes. 

Teacher educators, therefore, require guidance in facilitating PSTs' engagement with GenAI for content acquisition \cite{Marengo2024, Nyaaba2024Buddy}. With this, the role of the teacher educator evolves into that of a facilitator, moving away from traditional lecture-based teaching and creating more time for other activities. This approach aligns with the current teacher education curriculum in many developing countries, including Ghana, where the focus has shifted away from rigid lecture notes or pamphlets towards lesson plans and digital inquiry \cite{TTEL2017}.

\subsubsection{Promoting Pedagogical Modeling}
With GenAI handling content delivery, teacher educators can emphasize pedagogical modeling, a component often neglected in lecture-based systems. The potential of GenAI to support content knowledge also allows teacher educators to spend less time on direct content instruction and more time on pedagogical modeling \cite{ElSayary2024, Yu2023}. This is especially important in countries like Ghana, where the education system has historically been focused on lecture-based teaching, often at the expense of hands-on, practical training \cite{AddaeKyeremeh2024,AnnanBrew2022}. Practical activities such as collaborative learning, inquiry-based projects, and other pedagogical modeling as enshrined in the teacher education curriculum could be realized \cite{Kolb2014,Wu2024}. 

When teacher educators have the opportunity to model for PSTs how to engage students in active learning, critical thinking, and problem-solving activities, this will eventually prepare PSTs for the challenges of modern classrooms \cite{NTC2017}.

\subsubsection{Reducing Resistance to Change }
One of the significant challenges in teacher education is the persistence of dominant teaching practices, with educators often resisting pedagogical changes that affect curricula fidelity \cite{Jost2015}. GenAI has the potential to disrupt these entrenched practices by promoting content-based learning, which teacher educators tend to focus on heavily, often at the expense of practical elements like pedagogical modeling \cite{Coffie2019}. 

As GenAI supports content delivery, teacher educators will have little choice but to engage with other essential aspects of teaching, particularly those related to practical application and modeling, which have previously been neglected \cite{AddaeKyeremeh2024}. This transition diminishes the value of lecture-based instruction, encouraging educators to focus on pedagogical modeling and other neglected areas of teacher preparation. 

If students can easily acquire content knowledge outside of the classroom through GenAI, the focus of in-class teaching must shift towards practical and hands-on learning, such as pedagogical modeling. This will become a critical component of teacher education, as students will seek out classes not for content repetition, but for the skills and strategies they can only gain through active modeling and engagement.

\subsubsection{Facilitating Performance-Based Assessments}
Performance-based assessments are crucial for evaluating teaching effectiveness but are often overshadowed by standardized testing in developing countries \cite{Rimfeld2019}. GenAI supports this shift by enabling project-based assessments, portfolios, and presentations. Standardized testing has long dominated teacher education, significantly impacting pedagogical modeling and, by extension, general teaching practices at the K-12 level \cite{Angrist2004, Rimfeld2019}. 

Even with the current reforms in Ghana, because teaching heavily relies on lecture-based methods, assessment practices have shifted toward standardized tests rather than authentic, performance-based evaluations. With the influence of GenAI on content-based learning, teacher educators will no longer have the option to rely solely on lecture-driven teaching, pushing them toward a more practical and interactive approach to pedagogy. This shift will naturally lead to a rethinking of assessment strategies, moving away from standardized testing and toward performance-based assessments, as recommended in the curriculum. 

For example, teachers may focus more on project-based assessments, portfolios, and presentations, which align with student-centered, practical learning approaches \cite{NTC2017, TTEL2017}. This transformation will better prepare PSTs for the realities of modern classrooms, where practical application and performance are key to student success.

\subsubsection{Enhancing Resources Acquisition }
As teachers move away from delivering content that can easily be accessed through GenAI, there will be a need to enrich their teaching practices through practical and modeling approaches. As posited by \cite{Singh2022}, GenAI offers significant benefits in higher education by providing flexible learning opportunities for students and personalized feedback on assessments. This shift will potentially encourage teacher educators to actively seek out and incorporate diverse resources and materials that enhance learning in more dynamic ways. 

This is to say, teacher educators will be compelled to explore appropriate resources to support the interactive and experiential components of their classes, moving beyond just content delivery to creating more engaging and hands-on learning environments. Moreover, \cite{Li2024} study pointed out that GenAI can possibly assist teacher educators by automating and aligning Open Educational Resources (OER) for their teachings, reducing manual efforts in searching for teaching resources. With reference to GenAI reducing the burden of content delivery, teachers will no longer be able to avoid focusing on the practical aspects of their teaching, and the need for suitable teaching materials and resources will become more pronounced. 

This shift will push educators and institutions to address any gaps, ensuring that classrooms are well-equipped for modern, resource-rich teaching and learning experiences.

\subsubsection{Encouraging Digital Literacy and Critical Thinking}
Generative AI can play a crucial role in fostering digital literacy and critical thinking skills among PSTs. For instance, \cite{Zhang2024} highlights the transformative potential of GenAI in teacher education, particularly its role in improving digital literacy and critical thinking. As GenAI becomes more integrated into teacher education programs, it will encourage PSTs to engage with digital tools and resources, enhancing their ability to navigate and use technology effectively in their learning and teaching, which are often lacking in developing countries. 

Moreover, \cite{Suriano2025} demonstrats that PSTs' interaction with GenAI can enhance their complex critical thinking skills and problem-solving abilities. Furthermore, teacher educators can also use GenAI to create scenarios or problems that require PSTs to analyze, evaluate, and apply their knowledge to solve real-world challenges \citep{LiangWu2024}. In developing countries like Ghana, this role of GenAI will foster the digital literacy and critical skills components enshrined in the teacher education curriculum.

\subsubsection{Aligning Teaching with Industrial Needs }
A critical gap in education is the misalignment with industrial needs, as graduates often lack practical skills. The long-term impact of GenAI in transforming teacher education on the overall education system should not be underestimated. As PSTs become more confident in applying their knowledge and skills through pedagogical modeling and practical teaching, they will be more motivated to teach and engage their future students in meaningful ways, fostering teaching, learning, and skill development. This will potentially address key challenges like the lack of skilled personnel and unemployment present in current overemphasized content-based teaching \citep{Omoso2020}. 

For example, \cite{Quansah2019} study on the gap between academia and industry in Ghana highlights that most graduates felt inadequately prepared with practical skills, as their education focused primarily on memorization and passing exams rather than experiential learning. This shows the pressing need for a transformative teacher education program that will prioritize relevant, hands-on skills that meet job market demands.

\subsubsection{Supporting Teacher Professional Development }
Generative AI enhances teacher professional development (PD) by providing flexible, self-directed professional learning opportunities. Generative AI has the potential to significantly enhance teacher PD, fostering continuous growth and innovation in teaching methods. \citet{Siminto2023} study shows that GenAI can serve as a valuable tool in enhancing teacher educators' professionalism by offering quick access to educational resources, providing training recommendations, facilitating collaboration in teaching planning, and aiding teacher educators in accessing real-time information and planning. 

GenAI can help create a more self-directed PD model where teacher educators are not solely reliant on scheduled sessions but can engage with learning tools at their own pace \citep{Chen2024}. This can make PD more flexible and accessible, ensuring that teacher educators continuously update their skills and knowledge, staying current with the latest educational trends.

\section{Conclusion and Recommendations }
Even though research has shown that GenAI is currently transforming education in diverse ways, there is limited focus on how these GenAI can effectively enhance pedagogical practices or impact teacher education and education in general, particularly in developing countries. This study examined the transformative role that GenAI can play in teacher education within developing countries, particularly in contexts like Ghana, where teacher education has long faced challenges in implementing pedagogical modeling and performance-based assessments, despite these being encouraged by many developing countries in their teacher education curricula. The study argues that GenAI has the potential to support content knowledge acquisition, which currently occupies a large portion of teacher education programs, allowing teacher educators to divert their efforts toward pedagogical modeling and performance-based assessment. This shift will indirectly address several other challenges, such as resistance to change among teacher educators, lack of resources, overreliance on standardized testing and the practitioner-expertise gap in teacher education programs. These effects will set a chain reaction in motion, where PSTs will experience pedagogical modeling and pass on these practices to enhance K-12 learning outcomes. This, in turn, will help bridge the gap between education and the skills needed in the workforce, eventually reducing unemployment. However, there should be caution against the misuse of GenAI, as it has the potential to even undermine existing critical thinking and creativity skills that are nurtured through the current traditional teaching methods, like lecture-based content delivery and content-based assessments.  
To ensure responsible and effective integration, this study recommends urgent implementation of GenAI literacy in developing countries in a scaffolding approach. This would begin by helping PSTs understand GenAI's supportive role in education, followed by training in its ethical use, equipping them with prompt engineering skills, and enabling them to critically assess GenAI-generated content for biases and validity. Finally, transitioning PSTs into practical applications of GenAI to support their future students is essential. Additionally, empirical studies are recommended to explore the practical steps necessary to realize the integration and maximize the potential benefits of GenAI in teacher education.


\bibliography{sn-bibliography}

\end{document}